\documentclass[aps,pre,superscriptaddress,twocolumn,amsmath,amssymb,showpacs]{revtex4}
 \usepackage{graphicx}
\usepackage{dcolumn}
\usepackage{bm}
\begin{document}

 \title{Quantum Stochastic Synchronization}

 \author{Igor Goychuk}
   \affiliation{Institut f\"ur Physik,
   Universit\"at Augsburg,
   Universit\"atsstr. 1,
   D-86135 Augsburg, Germany}

 \author{Jes\'us Casado-Pascual}
 \author{ Manuel Morillo}
 \affiliation{F\'{\i}sica Te\'orica,
 Universidad de Sevilla, Apartado de Correos 1065, Sevilla 41080, Spain}


 \author{J\"org Lehmann}
 \affiliation{Department
 f\"ur Physik und Astronomie,
 Universit\"at Basel,
 Klingelbergstra\ss e~82,
 CH-4056~Basel, Switzerland}

 \author{Peter H\"anggi}
   \affiliation{Institut f\"ur Physik,
   Universit\"at Augsburg,
   Universit\"atsstr. 1,
   D-86135 Augsburg, Germany}

\date{\today}

\begin{abstract}
 We study within the spin-boson dynamics the synchronization of  quantum
 tunneling  with an external  periodic driving signal. As a main result
 we find that at  a sufficiently large system-bath coupling strength
 (Kondo parameter $\alpha>1$) the thermal noise  plays a constructive
 role in yielding both a frequency and a phase synchronization in a
 symmetric two-level system. Such riveting synchronization occurs when
 the driving frequency supersedes the zero temperature tunneling rate.
 As an application evidencing the effect, we consider a charge transfer
 dynamics in molecular complexes.
\end{abstract}


\maketitle

The phenomenon of synchronization of nonlinear classical systems with
external driving signals, possibly even in the  presence of randomness,
has increasingly been gaining importance and growing interest over the
last decade~\cite{book1,book2,review1,review2}. A particular intriguing
example is crowd synchrony when pedestrians fall into steps with the
intrinsic vibrations of a footbridge~\cite{Strogatz}. Physically related
to the synchronization phenomenon is the phenomenon of Stochastic
Resonance~\cite{sr} where the presence of noise can manifestly boost the
transduction of an information-carrying, weak signal. This effect is
enduring ongoing vitality in diverse fields that span from physics to
the life sciences. This fascinating phenomenon has notably been
generalized into the quantum regime only
recently~\cite{qsr1,qsr2,GrifoniHanggi98}, where its experimental
realization on the level of a nanomechanical quantum memory element  is
very feasible~\cite{mohanty}. The  extension of noise-induced
synchronization into the realm of quantum physics, however,  has not
been considered thus far. This latter task presents a challenge which,
apart from basic academic interest, also comprises  great potential for
nanoscience with promising applications ranging from quantum control to
quantum information processing.

With this work we undertake a first step in this direction. We consider
the  paradigm of a driven, symmetric quantum two-level  system that is
coupled to an Ohmic thermal bath with an exponential
cutoff~\cite{GrifoniHanggi98,GoychukHanggi05}. When operating within the
overdamped regime, we successfully reveal the existence of a
noise-assisted {\it quantum synchronization}. There are two relevant
energetic parameters characterizing the system-bath coupling: the
reorganization energy $\lambda$ and the cutoff energy $\hbar\omega_c$
which is associated to the spectral width of the medium modes which couple to
the tunneling particle. The two parameters can be combined into the
dimensionless coupling strength $\alpha=\lambda/(2\hbar\omega_c)$, i.e.,
the Kondo parameter. A third relevant energy scale constitutes the
tunneling energy $\hbar\Delta$. The most relevant regime for our
purposes corresponds to $\hbar\Delta \ll \hbar\omega_c, \lambda$ and
$\alpha>1$.

Dissipative quantum tunneling changes radically the physics of
synchronization. At zero temperature, the system can only tunnel towards
its lowest energy state when a biasing dc-signal is
applied. As the bias periodically changes its sign due to the action of
a driving field, tunneling makes the particle move periodically towards
its corresponding lowest energy state, as long as the driving period is
much longer than the typical tunneling time. Consequently, one  expects
that the system synchronizes with a driving periodic rectangular signal.
By contrast, in the absence of classical thermal noise, synchronization
in a overdamped bistable system driven by rectangular subthreshold
signals does not occur, as the overbarrier transitions do not exist.

Two interesting questions now emerge: What is the effect of quantum
noise at finite temperature for synchronization? How does the time scale
of the the external driving period impact this synchronization behavior?
Quantum bath fluctuations at finite temperature surely will disturb the
above mentioned perfect synchronization of the tunneling events as
imposed by the rocking driving field. At the same time, when the driving
period becomes shorter than the tunneling time, the zero temperature
synchronization will also be weakened. Nonetheless, it might very well
be possible that finite temperature quantum noise promotes and assists
synchronization when the driving period decreases. We demonstrate below
that this  indeed is the case.

To start, we consider the following archetype model of a dissipative,
driven two-level system (TLS)~\cite{leggett,WeissBook,GrifoniHanggi98,GoychukHanggi05}:
 \begin{eqnarray}
 \label{TLS}
 \hat H(t) &=& \frac{1}{2}\epsilon (t) \hat \sigma_z
 +\frac{1}{2}\hbar\Delta\hat \sigma_x + \frac{1}{2}\hat \sigma_z
 \sum_{j}\kappa_{j}(b^{\dagger}_{j}+ b_{j}) \nonumber \\ & &+
 \sum_{j}\hbar\omega_{j}(b^{\dagger}_{j} b_{j}+\frac{1}{2}) .
\end{eqnarray}
Herein, the operators ${\hat \sigma_{z}}$ and ${\hat\sigma_{x}}$ denote
the standard Pauli operators, $\epsilon (t)$ is a time-dependent energy
bias, and $\hbar\Delta$ in (\ref{TLS}) is the tunneling matrix element.
The boson operators $b^{+}_{j}$ and $ b_{j}$ correspond to normal mode
oscillators of the thermal bath
with frequencies $\omega_{j}$. The stochastic influence of the quantum
thermal bath is captured by an {\it operator} random force
$\hat\xi(t)=\sum_{j}\kappa_{j}(b^{\dagger}_{j} e^{i\omega_{j}t}+b_{j}
e^{-i\omega_{j}t})$.  It can be characterized by the spectral density
$J(\omega)=(\pi/\hbar)\sum_{j}^{}
\kappa_{j}^2\delta(\omega-\omega_{j})$. We assume that $J(\omega)$
acquires the Ohmic form with an exponential cutoff,
$J(\omega)=2\pi\hbar\alpha\omega e^{-\omega/\omega_c}$.  Due to the
Gaussian statistics of a harmonic bath, the statistical properties of
the quantum noise become completely determined by its equilibrium
autocorrelation function $\langle\hat{\xi}(t)\hat{\xi}(0)\rangle_{T}=
(\hbar/\pi)\int\limits_{0}^{\infty} J(\omega)[ \coth(\hbar\omega/2k_BT)
\cos(\omega t)-i\sin\omega t ] d\omega $. The driven spin-boson
Hamiltonian~\eqref{TLS} describes many situations, such as, e.g., the
electron transfer (ET) in a molecular dimer in azurin crystals~\cite{azurinPNAS}.
Then the low-frequency molecular vibrations provide
the bath and the time-dependent energy bias is given by $\epsilon(t)=r e
{\cal E}(t)$, where $r$ is the tunneling distance, $e$ denotes the
charge transferred, and  ${\cal E}(t)$ is the time-dependent, applied
electric field. 

In the
incoherent tunneling regime, the populations of the localized states
obey a nonstationary, Markovian dynamics.  In the presence of a
time-dependent bias, this description holds true for Ohmic friction at
an arbitrary temperature if the tunneling coupling remains small,
i.e. $\Delta\ll \omega_c$, and the coupling to the heat bath is
sufficiently strong, $\alpha>1/2$~\cite{WeissBook}.  The  populations
$p_{\pm}(t)=(1\pm\langle\sigma_z(t)\rangle_{T})/2$ then obey the balance
equations \cite{mastereq,GrifoniHanggi98,GoychukHanggi05}
\begin{equation}\label{balance}
 \dot{p}_{\beta}(t)=W_{-\beta}(t)p_{-\beta}(t)-W_{\beta}(t)p_{\beta}(t)
\end{equation}
with $\beta=\pm $, where the
time-dependent relaxation rates $W_{\beta}(t)$ are given by
\begin{eqnarray}\label{rate}
 W_{\pm }(t)&=&\frac{1}{2}\Delta^2\int_{0}^{\infty}
 d\tau\exp[-Q^{\prime}(\tau)]
 \nonumber\\&&\times\cos\left[Q^{\prime\prime}(\tau) \mp
 \frac{1}{\hbar}\int_{t-\tau}^{t} \epsilon(t')dt'\right]
\end{eqnarray}
within the Golden Rule approximation. The functions $Q^{\prime}(t)$
and $Q^{\prime\prime}(t)$ in (\ref{rate}) denote the real and
imaginary parts of 
\begin{equation}
 Q(t)= \frac{i\lambda t}{\hbar} +\frac{1}{\hbar^2}
 \int_{0}^{t}dt_1\int_{0}^{t_1}dt_2\,
 \langle \hat \xi(t_2)\hat \xi(0)\rangle_{T} \;,
\end{equation}
wherein $\lambda=\int_{0}^{\infty}d\omega
J(\omega)/(\pi\omega)$ is the bath reorganization
energy~\cite{mastereq}. Here,
$\lambda=2\alpha\hbar\omega_c$ and the function $Q(t)$ can be
evaluated in closed analytical form~\cite{WeissBook,OhmicExact}, reading
\begin{eqnarray}\label{ohmic}
 Q^{\prime}(t)&= &2\alpha \ln\Big \{\sqrt{1+\omega_c^2 t^2}
 \frac{\Gamma^2(1+\kappa)}
 {|\Gamma(1+\kappa+i\omega_{T} t)|^2}\Big\} \\
 Q^{\prime\prime}(t)&= &2\alpha\arctan(\omega_c t)\;.
\end{eqnarray}
In Eq.~(\ref{ohmic}), $\Gamma(z)$ denotes the Gamma function,
$\omega_{T}=k_BT/\hbar$, and $\kappa=\omega_{T}/\omega_c$. Note that in
the limit of an {\it adiabatic} driving varying on a time-scale ${\cal
T}$ such that $\omega_c{\cal T}, \alpha \omega_T{\cal T}\gg 1$, the
time-dependent transition rates $W_{\pm}(t)$ follow the {\it
instantaneous} value of the bias $\epsilon(t)$. In this limit, which is
assumed throughout in the following, the relaxation rates $W_{\pm}(t)$
obey the Boltzmann relation $ W_{+}(t)=e^{\epsilon(t)/k_BT}W_{-}(t)$.
Furthermore, in the high-temperature limit $k_BT\gg \hbar\omega_c$,
Eq.~(\ref{rate}) reduces to a generalized Marcus-Levich-Dogonadze form,
i.e., $W_{\pm}(t)=(\pi/2) \hbar \Delta^2/ \sqrt{4\pi\lambda
k_BT}\exp[-(\pm \epsilon(t) -\lambda)^2/4\lambda k_B T]$.  For $k_B T
\leq \hbar\omega_c$, explicit analytical expressions for the rates are
generally not available, except for $T=0$~\cite{WeissBook}; typically,
those must be determined numerically from Eq.~(\ref{rate}).

Using the Marcus-Levich-Dogonadze formula for an undriven molecular
system, one can estimate the  relevant parameter values. In particular,
for ET occurring in azurin dimer the values are $\lambda=0.25$ eV and
$\hbar\Delta=5\cdot 10^{-6}$ eV~\cite{azurinPNAS}. In molecular systems,
the cutoff frequency of low-frequency molecular vibrations ranges
between $\hbar\omega_c=5-20$ meV.  Choosing $\hbar\omega_c=12.5$ meV
yields $\alpha=10$ corresponding to a strongly
incoherent, overdamped tunneling dynamics.

It is worth noting that the present incoherent limit for the tunneling
dynamics of a driven, dissipative TLS allows for an effective {\it
quasiclassical} interpretation in terms of a classical random telegraph
process which is inhomogeneous in time. Its transition rates, however,
are governed by the manifestly {\it quantum expressions} detailed in
Eq.~(\ref{rate}). As such, our setup mimics the quantum analogue of a
noisy, classical synchronization behavior elaborated in
Refs.~\cite{Freund,Casado,Talkner}.  
A thought-experimental setup is that of a particle tunneling between
two localized states at random times subject to an external periodic
rectangular field with amplitude ${\cal E}_0$ and period ${\cal T}$
(frequency $\Omega=2 \pi /{\cal T}$).
One counts the number of jumps $N(t,t_0)$
within a time window $[t_0,t]$. Following
Refs.~\cite{Freund,Casado,Talkner}, we introduce the random phase
$\phi(t,t_0)=\pi N(t,t_0)$, which increases by $\pi$ at each switching
event (two subsequent switches correspond to a $2\pi$-cycle with random
duration). We define the mean frequency and diffusion coefficients
associated to the phase process as $\overline{
\Omega}_{\mathrm{ph}}:=\lim_{t\to \infty}
\langle\phi(t,t_0)\rangle/(t-t_0)$ and $2\overline{D}_{\mathrm{ph}}:=
\lim_{t\to \infty} \left[\langle \phi^2(t,t_0)\rangle -\langle
\phi(t,t_0)\rangle^2\right]/(t-t_0)$, respectively. To evaluate these
quantities, one can consider the joint probability $P_{\beta,n}(t)$ to
be in the state $\beta$ at time $t$ and to have $n$ jumps within the
time interval $[t_0,t]$. These probabilities can be obtained by
integrating the multi-time probability densities of the corresponding
stochastic trajectories~\cite{GH00}. They satisfy the normalization
condition $\sum_{n=0}^{\infty}\sum_{\beta=\pm}P_{\beta,n}(t)=1$ and
obey a master equation resembling Eq. (\ref{balance}):
\begin{eqnarray}\label{chain}
 \dot P_{\beta,n}(t)=
 W_{-\beta}(t)P_{-\beta,n-1}(t)-W_{\beta}(t)
 P_{\beta,n}(t)
\end{eqnarray}
for $n\geq 1$ and $\dot P_{\beta,0}(t)=-
W_{\beta}(t)P_{\beta,0}(t)$ for $n=0$. The populations of the states
can be obtained as $p_{\beta}(t)=
\sum_{n=0}^{\infty}P_{\beta,n}(t)$ and the probability
$P_n(t)$ to have $n$-jumps is then
$P_n(t)=P_{+,n}(t)+
P_{-,n}(t)$. Given $P_n(t)$, any moment
$\langle n^k(t)\rangle:=\sum_{n=0}^{\infty}n^kP_n(t)$ ($k=1,2,...$) can
be obtained.  Deriving explicit analytical expressions, however,
presents a  nontrivial task. Fortunately, the first two moments
are an exception to this rule. In particular, from  Eqs. (\ref{chain}) and
(\ref{balance}), it follows that the phase frequency
$\Omega_{\mathrm{ph}}(t):=\pi (d/dt)\langle n(t)\rangle$
can be expressed as
\begin{eqnarray}\label{freqold}
 \Omega_{\mathrm{ph}}(t)=\pi[W_{+}(t)p_{+}(t)+W_{-}(t)p_{-}(t)]\;.
\end{eqnarray}
This  result coincides with one  obtained earlier in Refs.
\cite{Talkner,Casado}. For the averaged phase we have $\langle
\phi(t,t_0)\rangle= \int_{t_0}^{t}\Omega_{\mathrm{ph}}(t')dt'$. To
obtain $\overline \Omega_{\mathrm{ph}}$, one must take the limit
$\overline \Omega_{\mathrm{ph}}=\lim_{t\to\infty} \langle
\phi(t,t_0)\rangle/(t-t_0)$. For a periodic driving the asymptotic
behaviors for both $p_{\beta}(t)$ and $\Omega_{\mathrm{ph}}(t)$ are
periodic functions of time with period $\cal T$. Thus,
$\overline{\Omega}_{\mathrm{ph}}=(1/{\cal T})\int_{0}^{{\cal T}}dt\,
\Omega_{\mathrm{ph}}^{(\infty)}(t)$, where
$\Omega_{\mathrm{ph}}^{(\infty)}(t)$ is given by Eq.~(\ref{freqold})
with $p_{\beta}(t)$ replaced by the asymptotic solution
$p_{\beta}^{(\infty)}(t)$ of the master equation (\ref{balance}), which
formally is obtained by letting $t_0\to -\infty$.

 The calculation of the phase diffusion coefficient $D_{\mathrm{ph}}(t):=
 (\pi^2/2)(d/dt)\left[\langle n^2(t)\rangle -\langle
 n(t)\rangle^2\right]$ in closed form turns out to be intricate and lengthy, yielding
 \begin{eqnarray}\label{D}
 2D_{\mathrm{ph}}(t)&=&\pi\Omega_{\mathrm{ph}}(t)-2\pi^2 \delta W(t) \sum_{\beta=\pm}
\beta \int_{t_0}^{t}dt'\,W_{\beta}(t')
  \nonumber \\
 &&\times p_{\beta}^2(t')
 \exp\left[-\int_{t'}^{t}W(\tau)d\tau \right],
 \end{eqnarray}
 where $\delta W(t)=W_{+}(t)-W_{-}(t)$  and $W(t)=W_{+}(t)+W_{-}(t)$.
 This remarkable {\it exact} result in Eq. (\ref{D}) was originally 
 derived in a different context
 in Ref. \cite{Casado} within a slightly different approach. Note
that  $D_{\mathrm{ph}}(t)$ still depends, like
$\Omega_{\mathrm{ph}}(t)$,
on the initial time $t_0$. This dependence is asymptotically 
lost in the limit
$t_0\to-\infty$, where $p_{\beta}(t)$ is replaced by the asymptotic
periodic solution $p_{\beta}^{(\infty)}(t)$. The evaluation
 of the mean diffusion coefficient $\overline D_{\mathrm{ph}}$ proceeds
 similarly to $\overline{\Omega}_{\mathrm{ph}}$.
 For the considered case of rectangular driving, 
 both, $\overline{\Omega}_{\mathrm{ph}}$ and
  $\overline D_{\mathrm{ph}}$ are evaluated to read:
 \begin{equation}
 \label{frequency}
 \overline\Omega_{\mathrm{ph}} = \frac {\pi W}{2} \left \{ 1- \delta
p_{\mathrm{eq}}^2 \left [ 1- \frac{4 \tanh \left (W {\cal T}/4 \right
)}{W {\cal T}}\right ]\right \},
 \end{equation}
 and
 \begin{eqnarray}\label{diffusion}
 2\overline D_{\mathrm{ph}} &=& \pi \,\overline \Omega_{\mathrm{ph}} -
 \frac{2 \pi^2}{\cal T} \delta p_{\mathrm{eq}}^4 \tanh^3\left(W {\cal
 T}/4\right) \nonumber\\ &&- \frac{\pi^2}{2\cal T}\delta
 p_{\mathrm{eq}}^2 \left(1- \delta
 p_{\mathrm{eq}}^2\right)\Big\{12\tanh\left(W {\cal T}/4\right)
 \nonumber\\ &&- W {\cal T} \left[ 1+2 \,{\rm sech}^2\left(W {\cal
 T}/4\right)\right] \Big\},
 \end{eqnarray}
where $W$ denotes the sum of the forward and backward rates in Eq.~(\ref{rate})
for a fixed value of the field ${\cal E}_0$, i.e., for $\epsilon(t)=\epsilon_0=er{\cal E}_0$, and $\delta
 p_{\mathrm{eq}}=\tanh(\epsilon_0/(2k_BT))$ is
 the absolute value of the difference of the
 equilibrium populations. The inverse
 Fano factor of the counting process $R:=\pi
 \overline\Omega_{\mathrm{ph}}/(2\overline D_{\mathrm{ph}})$ provides
 a reliable quality measure of synchronization~\cite{review1,review2}.

 The quantum features of the so derived synchronization quantities are rooted in the quantum
 rate expressions entering Eqs.~(\ref{frequency}) and
 (\ref{diffusion}). By analogy with the findings for quantum stochastic resonance (QSR) in
 symmetric quantum systems \cite{qsr1}, one could expect that there is
 no thermal noise assisted synchronization for $\alpha\leq 1$. Indeed,
 we could not identify {\it noise-assisted} synchronization in this
 parameter regime. In contrast, employing the reasoning put forward for
 QSR within $\alpha>1$ in Ref.~\cite{qsr2},  one
 then supposes  quantum synchronization to emerge in this latter regime. 
 
\begin{figure*}[ht]
 \includegraphics[width=14cm]{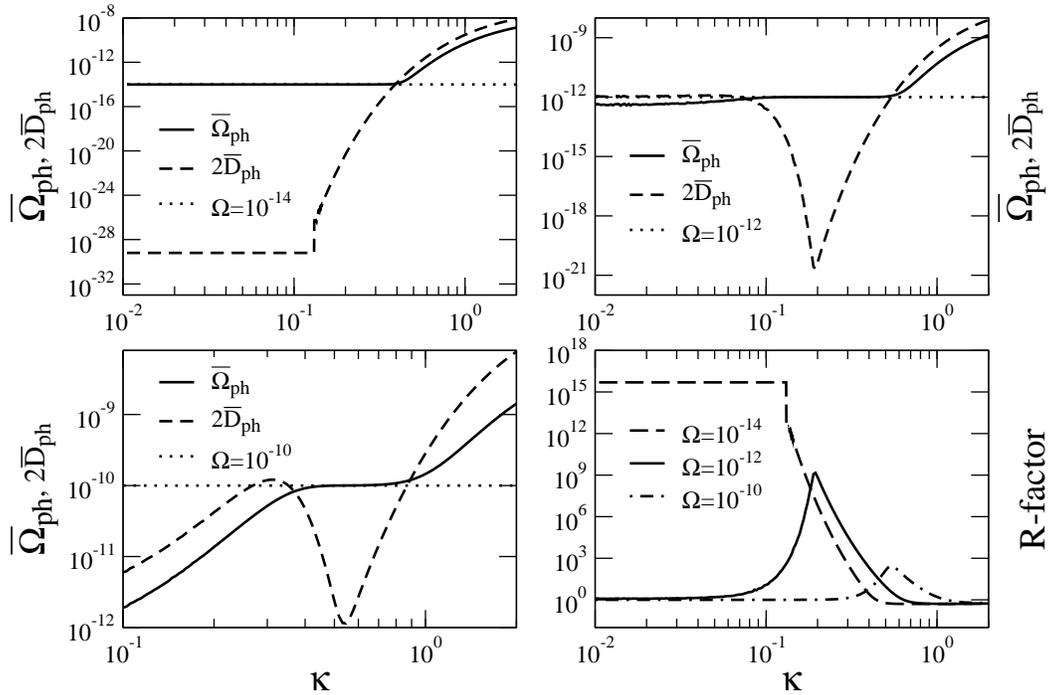}
 \caption{\label{fig1} Mean jump frequency $\overline\Omega_{\rm ph}$,
   phase diffusion coefficient $\overline D_{\rm ph}$ (upper right, upper left
   and bottom left panels) and the synchronization quality factor $R$ (bottom
   right panel) versus the scaled temperature $\kappa=k_BT/\hbar\omega_c$ for
   three values of the driving frequency $\Omega$. The parameter values used
   are: $\alpha=10$, $\Delta=4\cdot 10^{-4}$, $\epsilon_0=5$. Frequencies are
   scaled with $\omega_c$ and energies with $\hbar \omega_c$. For
   $\hbar\omega_c=12.5$ meV, $\omega_c\approx 1.9\cdot 10^{13}$ 1/sec, and
   $r=14.9$ \AA $\,$ (taken to match the charge transfer in molecules
   from Ref.~\cite{azurinPNAS})
   yields ${\cal E}_0\approx
   4.19\cdot 10^{4}$ V/cm. The value $\kappa=1$ corresponds approximately to
   $145$ K.}
 \end{figure*}

Indeed, for $\alpha>1$ the synchronization scenario depends sensitively on the
value of the driving frequency. As discussed above, for driving time scales with a
frequency much smaller than the tunneling rate at $T=0$, i.e., $\Omega
\ll W_{T=0}=\pi \Delta^2 [2 \omega_c \Gamma(2 \alpha)]^{-1}
[\epsilon_0/(\hbar\omega_c)]^{2\alpha-1} \exp[-\epsilon_0/
(\hbar\omega_c)]$, $\overline{\Omega}_{\mathrm {ph}}$ matches
 the external frequency for sufficiently low temperatures. This is depicted in the upper left
panel of Fig.~\ref{fig1}, where in addition the behavior of
$\overline{D}_{\mathrm{ph}}$ is presented. Notice that quantum
synchronization at sufficiently low temperature indeed is impressively well accomplished, as it is
reflected by the  large values of the $R$-factor (see the lower
right panel of Fig.~\ref{fig1}). As the temperature increases above a
certain critical-like value, however, the $R$-factor diminishes and the quality
of synchronization deteriorates.

With increasing driving frequency such that the condition $\Omega \ll W_{T=0}$
is no longer being  met, we enter the regime of thermally induced phase
synchronization in the quantum regime, as depicted in the upper right
and lower left panels in Fig.~\ref{fig1}. We observe a range of
temperature where the external frequency and
$\overline{\Omega}_{\mathrm{ph}}$ coincide. Moreover, the diffusion
coefficient exhibits a pronounced cusp-like minimum. The results plotted
in those two panels convincingly illustrate the constructive role of
quantum thermal noise for quantum
synchronization. Notice that with increasing driving frequency the
quality factor associated to the synchronization effect starts to diminish,
 see the lower right panel in Fig.~\ref{fig1}. Moreover, the temperature
range where frequency locking is observed shrinks. The quality factor
$R$ displays a cusp-like feature similar to the one discussed recently  by Park
and Lai~\cite{Park} within the context of noisy, classical synchronization.
Upon further increasing  the   driving frequency, the tunneling rate is too
slow compared to the driving frequency for the tunneling  dynamics to
follow the external oscillations. Thus, quantum synchronization is lost 
(not depicted).

In conclusion, we have discovered the existence of a quantum stochastic 
synchronization
in an externally driven spin-boson system which undergoes tunneling
transitions between two states.  We exemplified the phenomenon for
non-adiabatic charge transfer in molecular complexes. 
The estimates of parameter values for an experimental test of
our theoretical predictions are: driving frequencies
$\Omega
\sim 10^{-1}-10^{3}$ s$^{-1}$, electrical field strength ${\cal E}\sim
10^{4}-10^{5}$ V/cm, and temperatures $T\sim 20-100$ K, 
which are readily achieved
in the laboratory.  The main quantum features of the
discussed synchronization  phenomenon are robust and they are not
critically dependent on  the details of the underlying dissipation
mechanism. This is so because the phenomenon of quantum synchronization
is primarily based on the existence of a
low-temperature limit of the tunneling rates. 
We, consequently,
expect our very general results to be useful in other contexts
such as for optimizing the noisy dynamics of nano-mechanical systems in
the quantum regime~\cite{mohanty}.
The authors are confident that these riveting findings
for quantum synchronization will spur
experimental interests for diverse systems that involve a controllable quantum tunneling
between distinct quantum states.

{\it Acknowledgements.}
  J. C.-P. and M. M. acknowledge the support of the Ministerio de Educaci\'on y
  Ciencia of Spain (FIS2005-02884) and the Junta de Andaluc\'{\i}a. I. G.
  and P. H. acknowledge the support by the DFG through SFB 486.
  We also acknowledge ESF-support via the programme ``Stochdyn''.

 \end{document}